\documentclass[twoside,aps,prl,twocolumn]{revtex4-1}
\usepackage[T1]{fontenc}
\usepackage[latin9]{inputenc}
\usepackage{units}
\usepackage{amstext}
\usepackage{amssymb}
\usepackage{graphicx}

\makeatletter

\newcommand{\lyxmathsym}[1]{\ifmmode\begingroup\def\b@ld{bold}
  \text{\ifx\math@version\b@ld\bfseries\fi#1}\endgroup\else#1\fi}

\providecommand{\tabularnewline}{\\}

\@ifundefined{textcolor}{}
{%
 \definecolor{BLACK}{gray}{0}
 \definecolor{WHITE}{gray}{1}
 \definecolor{RED}{rgb}{1,0,0}
 \definecolor{GREEN}{rgb}{0,1,0}
 \definecolor{BLUE}{rgb}{0,0,1}
 \definecolor{CYAN}{cmyk}{1,0,0,0}
 \definecolor{MAGENTA}{cmyk}{0,1,0,0}
 \definecolor{YELLOW}{cmyk}{0,0,1,0}
}

\makeatother

\begin{document}

\title{Efficient Spin Injection into Silicon and the Role of the Schottky
Barrier}

\author{André Dankert$^{*}$}

\email{andre.dankert@chalmers.se}

\affiliation{Chalmers University of Technology, Department of Microtechnology
and Nanoscience, Quantum Device Laboratory; Göteborg, Sweden}

\author{Ravi S. Dulal}

\affiliation{Chalmers University of Technology, Department of Microtechnology
and Nanoscience, Quantum Device Laboratory; Göteborg, Sweden}

\author{Saroj P. Dash$^{**}$}

\email{saroj.dash@chalmers.se}

\affiliation{Chalmers University of Technology, Department of Microtechnology
and Nanoscience, Quantum Device Laboratory; Göteborg, Sweden}

\begin{abstract}
Implementing spin functionalities in Si, and understanding the fundamental
processes of spin injection and detection, are the main challenges
in spintronics. Here we demonstrate large spin polarizations at room
temperature, 34\% in n-type and 10\% in p-type degenerate Si bands,
using a narrow Schottky and a SiO$_{2}$ tunnel barrier in a direct
tunneling regime. Furthermore, by increasing the width of the Schottky
barrier in non-degenerate p-type Si, we observed a systematic sign
reversal of the Hanle signal in the low bias regime. This dramatic
change in the spin injection and detection processes with increased
Schottky barrier resistance may be due to a decoupling of the spins
in the interface states from the bulk band of Si, yielding a transition
from a direct to a localized state assisted tunneling. Our study provides
a deeper insight into the spin transport phenomenon, which should
be considered for electrical spin injection into any semiconductor. 
\end{abstract}
\maketitle
Spintronics exploits the spin of the electron, rather than its charge,
for information storage and processing.\cite{Zutic2004,Chappert2007,Awschalom2007}
Developing methods for efficient injection, controlled manipulation,
and sensitive detection of electron spins in semiconductors has the
potential to profoundly affect information technology. The strong
interest in silicon spintronics rises from the expected long spin
coherence length and its industrial dominance.\cite{Jansen2012a}
Creating spin polarized carriers in Si by using polarized light\cite{Lampel1968a},
hot electrons spin injection\cite{Appelbaum2007}, tunnel spin injection\cite{Dash2009,Jonker2007,Li2009,Suzuki2011,VantErve2007,Li2011,Jeon2011,Jansen2010,Jansen2010a},
Seebeck spin tunneling\cite{LeBreton2011a}, and dynamical spin pumping
methods has been demonstrated recently\cite{Shikoh2013a}. The use of ferromagnetic tunnel contacts to inject and detect spin
polarizations in Si has been recognized as the most viable and robust
method among them.\cite{Jansen2012c,Jansen2012a} Recently, optical
detection of spin polarization, through analysis of the degree of
polarization of the light emitted by spin\textendash{}LED structures,
showed 30\% spin polarization in Si at 77K.\cite{Jonker2007,Li2009}
In comparison, an all-electrical nonlocal measurement method that
represents the detection of spin accumulation in the Si gives rise
to a spin polarization of less than 1\% \cite{Suzuki2011,VantErve2007}.
More recently, it has become possible to probe large spin accumulations
directly underneath the injection contact up to room temperature by
means of the Hanle effect, using a three-terminal configuration \cite{Dash2009,Li2011,Jeon2011,Jansen2012a}.
In these experiments spin-polarization values of 5\% in n-type Si
at $\unit[300]{K}$ have been observed by using Al$_{2}$O$_{3}$,
MgO, plasma oxidized SiO$_{2}$, and graphene tunnel barriers together
with ferromagnetic contacts.\cite{Jansen2012a,Li2011,Van'tErve2012,Dash2009,Jeon2011}
These low values of spin-polarization obtained through electrical
methods are mainly due to the lack of high-quality tunnel barriers
on Si, which prevents efficient spin injection and detection. The
challenge is to achieve spin polarization in Si close to the tunnel
spin-polarization values of ferromagnetic contacts.

Another important issue is the mechanism of spin injection and detection
in semiconductors. The semiconductor/tunnel-barrier/ferromagnet contacts
are associated with a Schottky barrier and carrier depletion at the
semiconductor surface. This gives rise to different spin-transport
processes, depending on the profile of the Schottky barrier \cite{Jansen2007,Dery2007a}.
It has been proposed that, for degenerate semiconductors, a very narrow
Schottky barrier would allow direct spin-polarized tunneling, while
in nondegenerate semiconductors, the presence of a wider Schottky
barrier would be expected to change the transport mechanism \cite{Jansen2007}.
In the latter case, the depletion region is too wide for tunneling
and thermionic and localized state-assisted transport are expected
to dominate. It is proposed that such transport processes give rise
to anomalous behavior in spin accumulation and detection \cite{Jansen2007,Tran2009,Dery2007a,Jansen2012b}.
Although anomalous changes in the sign of spin signals have been observed
in different semiconductors \cite{Salis2011,Lou2007,Jeon2012b}, experiments
to elucidate this effect have not yet been performed.

In this article, we demonstrate the creation of large spin polarizations
in n-type and p-type Si, using ozone oxidized SiO$_{2}$ as a tunnel
barrier, and we address the role played by the profile of the Schottky
barrier in the processes of spin injection and detection. Degenerate
Si provides a very narrow Schottky barrier, which allows efficient
spin injection and detection in the direct-tunneling regime over a
wide temperature range. Furthermore, increasing the width of the Schottky
barrier in nondegenerate Si results in an anomalous sign change of
the spin signal in the low bias regime.  This can be due to the change
in transport processes across the interface from direct to indirect
tunneling, since spins accumulated in localized states can be decoupled
from the Si bands by the Schottky barrier. Our observations are generic
in nature, and are also valid for different tunnel barrier materials
on Si.

\part*{Results}

\begin{figure*}
\begin{centering}
\includegraphics{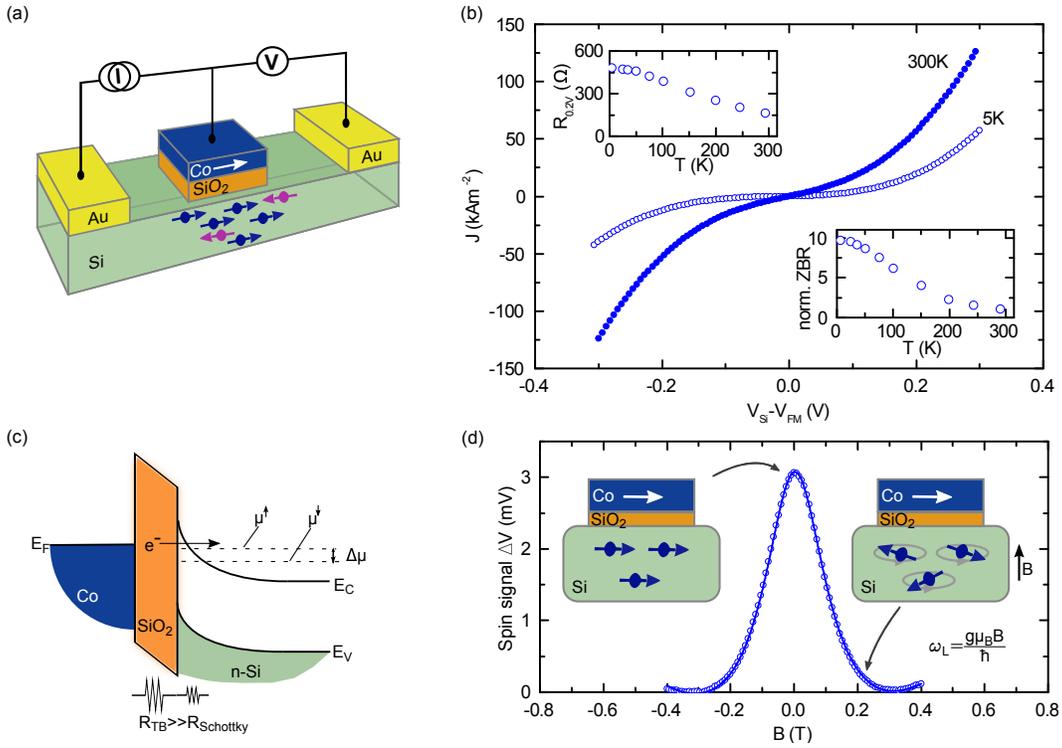}
\end{centering}

\caption{\textbf{Large spin signal in degenerate n-type Si:}{\footnotesize{
}}(a) Three-terminal device geometry for injection and detection of
spin polarization in Si with SiO$_{2}$/Co tunnel contacts. (b) Current
density versus bias voltage of the n++ Si/SiO$_{2}$/Co tunnel contact
measured in a three-terminal geometry at different temperatures. Insets:
Temperature dependence of junction resistance at bias voltages of
zero and $\unit[+200]{mV}$. (c) Energy-band diagram showing the injection
of spin-polarized current through a ferromagnetic tunnel contact into
n-type Si, creating a majority spin accumulation and spin-splitting
of electrochemical potential in the conduction band.. (d) Electrical
detection of large spin polarization in n-type Si at $\unit[300]{K}$
through the Hanle effect. The spin signal of $\unit[3]{mV}$ is observed
for $\unit[+1]{V}$ bias voltage and $\unit[+34]{mA}$ bias current.
The solid line is a Lorentzian fit with a lower limit for the spin
lifetime $\tau_{eff}=\unit[50]{ps}$. Left panel: The maximum spin
accumulation in the absence of external magnetic field B. Right panel:
A finite perpendicular magnetic field causes spin precession at the
Larmor frequency, and results in the suppression of the spin accumulation.}
\label{nppSi}

\end{figure*}
\textbf{Large electron spin polarization in degenerate n-type Silicon.}
To demonstrate large spin accumulations by direct tunneling, we used
SiO$_{2}$/Co tunnel contacts on degenerate n-type Si (n++ Si; measured
electron density $\unit[n=3\cdot10^{19}]{cm^{-3}}$ at $\unit[300]{K}$).
The SiO$_{2}$ barrier was prepared by ozone oxidation (see Methods).
Electrical measurements were performed in a three-terminal geometry
(Fig. \ref{nppSi}a), in which the same tunnel interface is used for
injection and for detection of spin accumulation in Si \cite{Lou2006,Dash2009,Tran2009}.{\footnotesize{
}}
\begin{figure*}[t]
\begin{centering}
\includegraphics[width=14.8cm]{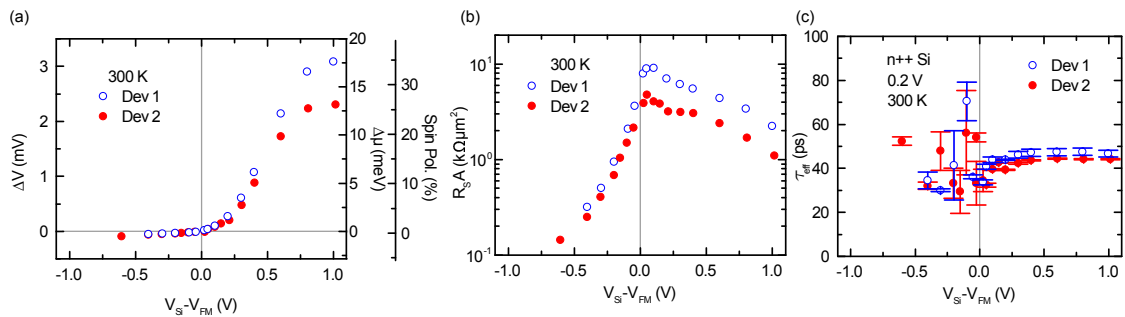} 
\end{centering}

\caption{\textbf{Bias dependence of the spin signal for degenerate n-type Si:}
Measurements are shown for two different devices. (a) The bias dependence
of the measured Hanle voltage signal ($\Delta V$), the calculated
spin splitting ($\Delta\mu$), and the spin polarization ($P$) created
in the Si conduction band due to spin injection at $\unit[300]{K}$.
The bias voltages $V_{Si}-V_{FM}>0$ and $V_{Si}-V_{FM}<0$ correspond
to spin injection and spin extraction, respectively. (b) Bias dependence
of spin-RA product ($\frac{\Delta V}{J}$) at $\unit[300]{K}$. (c)
The bias dependence of the effective spin lifetime ($\tau_{eff}$)
at $\unit[300]{K}$ with error bars.}

\label{nppSiDat-Vdep} 
\end{figure*}

\begin{figure*}[t]
\begin{centering}
\includegraphics[width=14.8cm]{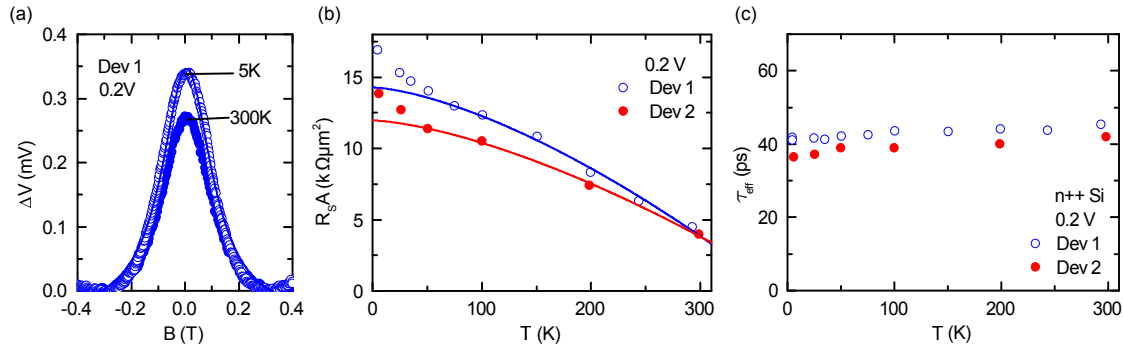} 
\end{centering}

\caption{\textbf{Temperature dependence of the spin signal for degenerate n-type
Si:} Measurements are shown for two different devices. (a) Hanle spin
signal at $\unit[5]{K}$ and $\unit[300]{K}$ for an applied bias
voltage of $\unit[+0.2]{V}$ (b) Temperature dependence of spin-RA
at an applied bias voltage of $\unit[+0.2]{V}$. The lines represent
the theoretical prediction for spin accumulations created through
direct tunneling: $R_{S}A\propto1-\alpha T^{\frac{3}{2}}$.\cite{Shang1998}
(c) Temperature dependence of the effective spin lifetime (error bars
are smaller than data points).}

\label{nppSiDat-Tdep} 
\end{figure*}

The contacts show the tunneling behavior characterized by nonlinear,
quasisymmetric $J\lyxmathsym{\textendash}V$ characteristics with
a weak temperature dependence (Fig. \ref{nppSi}b)\cite{Jonsson-Akerman2000}.
We obtain a resistance-area-product in the junction of $R_{junc}A\sim\unit[4]{\Omega mm^{2}}$
at room temperature, which increases only by factor two when cooling
down to $\unit[5]{K}$. Additional to the low temperature dependence,
we observed an exponential dependence of junction resistance with
SiO$_{2}$ tunnel barrier thickness (supplementary Fig. S1). These
results indicate the growth of an uniform and pinhole free SiO$_{2}$
tunnel barrier on Si. Such barriers were used in ferromagnetic tunnel
contacts of Co/SiO$_{2}$ to create an majority spin accumulation
and splitting of electrochemical potential in the conduction band
of n++ Si (Fig. \ref{nppSi}c). The Hanle effect is used to control
the reduction of the induced spin accumulation by applying an external
magnetic field ($B$) perpendicular to the carrier spins in the Si.
The spin accumulation decays as a function of $B$ with an approximately
Lorentzian line shape given by $\Delta\mu\left(B\right)=\frac{\Delta\mu\left(0\right)}{1+\left(\omega\tau_{S}\right)^{2}}$,
where $\Delta\mu\left(0\right)$ and $\Delta\mu\left(B\right)$ are
respectively the spin accumulation in zero magnetic field and in a
finite perpendicular magnetic field, and $\tau_{S}$ is the spin lifetime
\cite{Dash2009}. Figure \ref{nppSi}d shows the measurement of a
large electrical Hanle signal, $\Delta V=\unit[3]{mV}$, for a Co/SiO$_{2}$/n++
Si junction at 300 K and for a constant tunnel current of $I=\unit[34]{mA}$
at a bias voltage of $\unit[1]{V}$. In the linear response regime,
this large signal corresponds to a spin splitting of $\Delta\mu=\frac{2\Delta V}{\text{TSP}}=\unit[20]{meV}$
(with assumed SiO$_{2}$/Co contact tunnel spin polarization TSP$=0.35$).
The Fermi\textendash{}Dirac distributions for the majority ($n{}^{\uparrow}$)
and minority ($n^{\downarrow}$) electron spin densities are found
to be $\unit[2\cdot10^{19}]{cm^{-3}}$ and $\unit[10^{19}]{cm^{-3}}$,
respectively \cite{Dash2009,Jansen2012a}. This corresponds to a giant
electron spin polarization of $P=\frac{n^{\uparrow}-n^{\downarrow}}{n^{\uparrow}+n^{\downarrow}}=34\%$
in the n++ Si conduction band, which is nearly one order of magnitude
greater than previously reported values\cite{Jansen2012a,Li2011,Van'tErve2012,Dash2009,Jeon2011}
and close to the TSP of the ferromagnetic tunnel contact at room temperature.
The low-defect density of the ozone-oxidized SiO$_{2}$ tunnel barrier
and its better interface quality with Si offers a great advantage,
suggesting a means to produce a large spin polarization in Si at room
temperature.

Investigating the bias voltage and temperature dependence of the Hanle
spin signal provides more detailed quantitative information about
spin-tunnel processes. The bias dependence of the Hanle spin signal
($\Delta V$), the calculated spin splitting and spin polarization
are presented in Fig. \ref{nppSiDat-Vdep}a. For $V_{Si}-V_{FM}>0$
(reverse bias), the spin injection from the SiO$_{2}$/Co tunnel contact
creates a majority spin accumulation, while for $V_{Si}-V_{FM}<0$
(forward bias), the spin extraction creates a minority spin accumulation
in the Si. The magnitude of the spin accumulation varies asymmetrically
with the bias voltage across the junction, even though the current
density is almost symmetric with respect to bias-voltage polarity.
The measured Hanle signal for the majority and minority spin accumulation
are opposite in sign. For spin injection, the signal increases linearly
with low bias current, and saturates at higher bias (see supplementary
Fig.~S10 for bias current dependence). The Hanle curves at different
bias voltages are presented in the supplementary information (supplementary
Fig.~S2).
\begin{figure*}[!t]
\begin{centering}
\includegraphics[width=14.8cm]{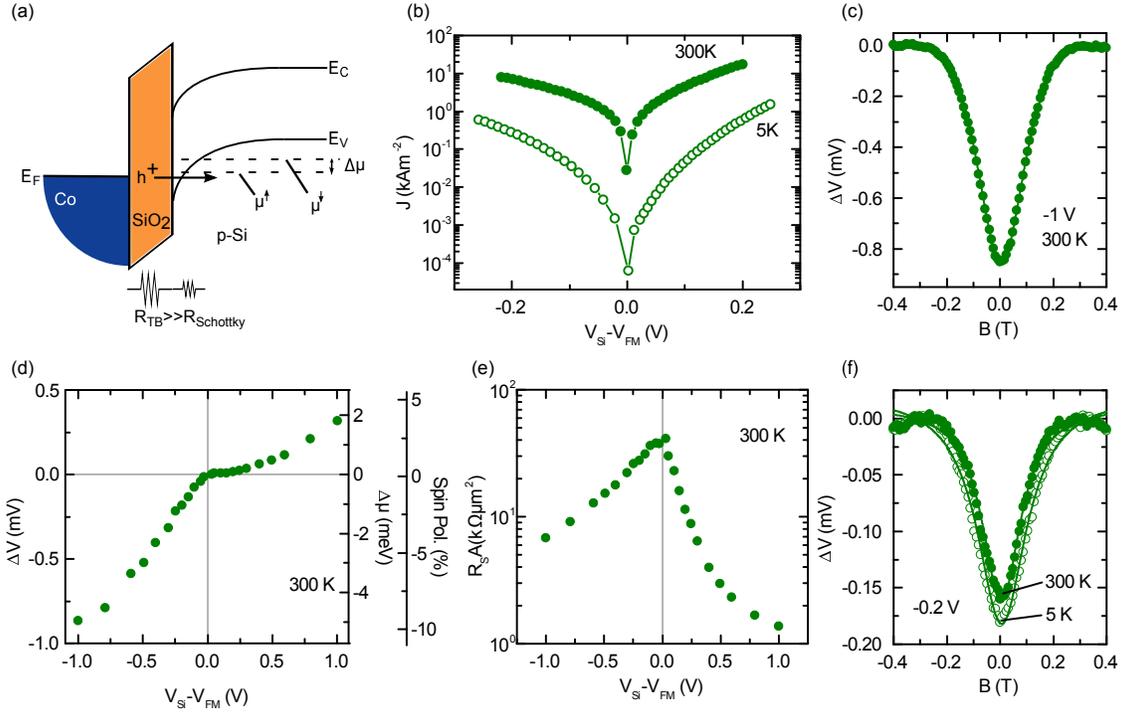} 
\end{centering}

\caption{\textbf{Spin signal for degenerate p-type Si:}\textsf{ }(a) Energy-band
diagram for SiO$_{2}$/Co tunnel contact with p++ Si. The injection
of spin-polarized holes creates a spin accumulation and spin-splitting
of electrochemical potential in the Si valence band. (b) Current density\textendash{}bias
voltage characteristics at 5 and $\unit[300]{K}$. The bias voltages,
$V_{Si}-V_{FM}<0$ and $V_{Si}-V_{FM}>0$, correspond to spin injection
and spin extraction, respectively. (c) Large Hanle signal measured
for spin-polarized hole injection at $\unit[-1]{V}$ bias voltage
and $\unit[-3.17]{mA}$ bias current at $\unit[300]{K}$. (d) Bias
dependence of Hanle spin signal at $\unit[300]{K}$, calculated spin
splitting and spin polarization created in the Si valence band. (e)
Bias dependence of $R_{S}A$ at $\unit[300]{K}$. (f) Temperature
dependence of Hanle spin signal measured at 5 and $\unit[300]{K}$
at a bias voltage of $\unit[-0.2]{V}$.}

\label{PppDat} 
\end{figure*}

The spin-resistance-area product ($R_{S}A=\frac{\Delta V}{J}$) is
found to be in the range $\unit[1-4]{k\Omega\mu m^{2}}$ (Fig. \ref{nppSiDat-Vdep}b),
which is large, compared to theoretical predictions \cite{Fert2001,Jansen2012a}.
In the diffusive regime, $R_{S}A$ should be equal to $P^{2}\rho_{Si}L_{sd}=\unit[10]{\Omega\mu m^{2}}$,
where $\rho_{Si}=\unit[3]{m\Omega cm}$ at $\unit[300]{K}$, and $L_{sd}$
is the spin diffusion length \cite{Jansen2012a}. Although the experimental
values are large, we can rule out any enhancement of the spin signal
by tunneling through localized states over the full temperature range.
The weak temperature dependence of the Hanle spin signal (Fig. \ref{nppSiDat-Tdep}a)
and $R_{S}A$ (Fig. \ref{nppSiDat-Tdep}b) at low bias voltages matches
the theoretical predictions $R_{S}A\propto1-\alpha T^{\frac{3}{2}}$.\cite{Dash2009,Shang1998}
 This indicates a true spin accumulation in the Si conduction band
over the full temperature range, since localized interface states
are expected to provide a larger temperature dependence of $R_{S}A$
\cite{Tran2009,Jain2012b}. The thinner, low resistance Schottky barrier
of this highly doped n++ Si devices couples well the localized states
and the Si conduction band allowing for direct spin-polarized tunneling
\cite{Jansen2012b,Jansen2010}. It should also be noted that the weak
temperature dependence of SiO$_{2}$ has an advantage over other oxide
tunnel barriers such as Al$_{2}$O$_{3}$\cite{Jansen2012a} and MgO\cite{Jeon2011,Jain2012b},
in which $R_{S}A$ increases exponentially at lower temperatures.
Such low temperature dependence of spin signal has also been reported
using plasma SiO$_{2}$ tunnel barriers on Si \cite{Li2011}. From
the separately measured diffusion constant $D=\unit[2.9]{\frac{cm^{2}}{s}}$
and the Lorentzian fitting of the Hanle curves, the lower limit for
the spin lifetime and the spin diffusion length are found to be $\tau_{eff}=\frac{\hbar}{g\mu_{\text{B}}\Delta B}=\unit[50]{ps}$
and $\lambda_{sf}=\sqrt{D\tau_{eff}}=\unit[120]{nm}$, respectively,
at room temperature. These values match very well previously reports
for such highly doped Si at room temperature \cite{Jansen2012c,Dash2009,Dash2011a}.
Furthermore, the spin lifetime is found to be independent of both
temperature and bias voltage (Fig. \ref{nppSiDat-Vdep}c and \ref{nppSiDat-Tdep}c),
supporting the direct tunneling and detection of spin accumulations
in the bulk Si conduction band over the measured temperature and bias
voltage ranges.

\noindent \textbf{Large hole spin polarization in degenerate p-type
Silicon.} By employing similarly fabricated SiO$_{2}$/Co tunnel contacts
on degenerate p-type Si (p++ Si, hole density $p=1.8\cdot\unit[10^{19}]{cm^{-3}}$
at $\unit[300]{K}$), we have studied spin-polarized hole accumulations
through direct tunneling. Figure \ref{PppDat}a shows the energy-band
diagram and the creation of spin splitting in the valence band of
p++ Si due to electrical spin injection. The $J-V$ characteristics
of the tunnel junction at $\unit[5]{K}$ and $\unit[300]{K}$ are
shown in Fig. \ref{PppDat}b. Hanle measurements at room temperature
demonstrate a large spin signal of $-\unit[0.87]{mV}$ at $\unit[-1]{V}$
bias voltage and $\unit[-3.17]{mA}$ bias current (Fig. \ref{PppDat}c),
which corresponds to a spin splitting in the p++ Si valence band of
$\Delta\mu=\frac{2\Delta V}{\text{TSP}}=\unit[5]{meV}$, and majority
($p^{\uparrow}$) and minority ($p^{\downarrow}$) hole spin densities
of $\unit[0.99\cdot10^{19}]{cm^{-3}}$ and $\unit[0.81\cdot10^{19}]{cm^{-3}}$,
respectively. This indicates that hole spin polarization is large,
at $P=\frac{p^{\uparrow}-p^{\downarrow}}{p^{\uparrow}+p^{\downarrow}}=10\%$.
It should be noted that we have not taken into consideration the valence-band
spin splitting that would be present in p-type Si due to the spin-orbit
interaction.

\begin{figure*}[t]
\begin{centering}
\includegraphics{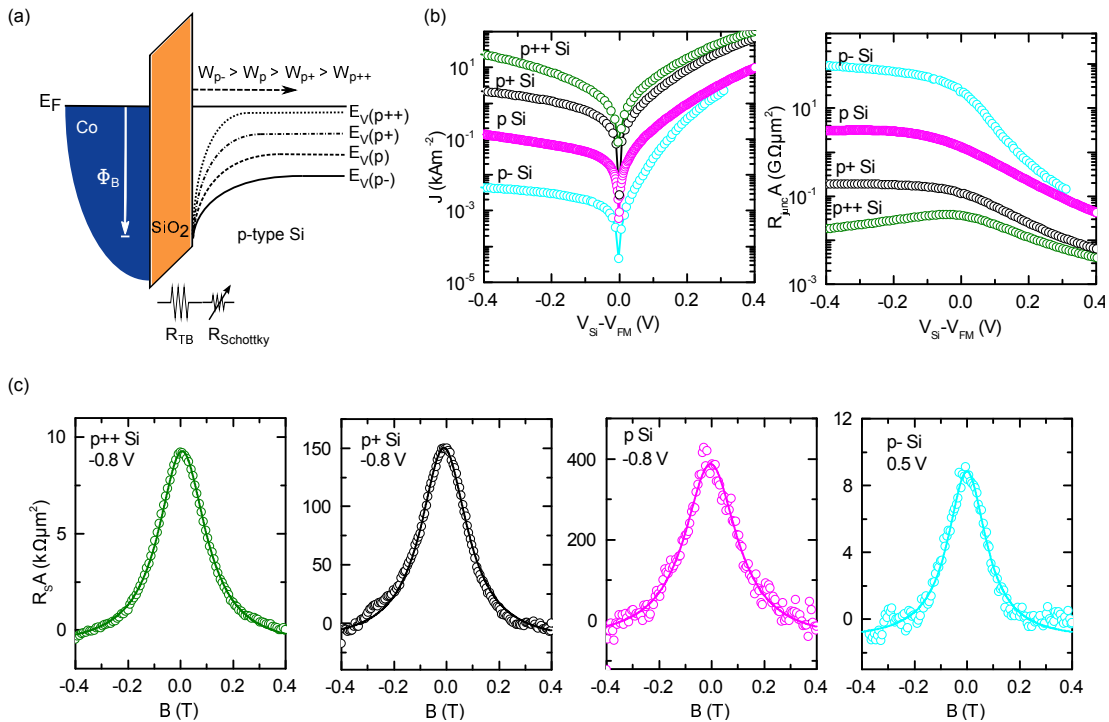} 
\end{centering}

\caption{\textbf{Tailored Schottky barrier width and Hanle spin signals for
p-type Si:} Si devices with four different boron doping concentrations
were studied. (a) Energy-band diagram for p-type Si/SiO$_{2}$/Co
showing doping-dependent Schottky barrier width. (b) Bias voltage
dependence of current density and $R_{junc}A$ at $\unit[300]{K}$.
(c) Hanle spin signals measured at $\unit[300]{K}$.}

\label{pTypeIntro} 
\end{figure*}

The bias dependence of the Hanle spin signal is shown in Fig. \ref{PppDat}d.
Spin injection ($V_{Si}-V_{FM}<0$) and spin extraction ($V\mbox{\ensuremath{_{Si}}}-V_{FM}>0$)
produce a net excess of holes, with majority and minority spin, respectively.
As expected, a change in the sign of the Hanle signal is observed,
corresponding to these majority and minority spin accumulations. The
Hanle signals are presented in supplementary Fig. S3, as measured
at different bias voltages. The $R_{S}A$ shows asymmetric behavior,
decreasing more quickly for the case of spin extraction than in the
case of spin injection (Fig. \ref{PppDat}e). Although the $R_{S}A$
values observed are larger than expected on the basis of the spin-diffusion
model \cite{Fert2001,Jansen2012a,Tran2009}, the narrow Schottky barrier
for p++ Si should rule out spin accumulation in localized states \cite{Jain2012b}.
This is supported by the very weak temperature dependence of the Hanle
signal observed at a low bias voltage of $\unit[-200]{mV}$, providing
a clear indication of the true spin accumulation in the Si valence
band (Fig. \ref{PppDat}f). From the Lorentzian fitting of the Hanle
curves, the lower limit for the spin lifetime is found to be $\tau_{S}\sim\unit[50]{ps}$,
and spin diffusion length is seen to be $L_{SD}\unit[\sim80]{nm}$.
These values are similar to those from previous reports on degenerate
p-type Si as measured by electrical spin injection \cite{Dash2009},
thermal spin injection \cite{LeBreton2011a}, and spin pumping \cite{Shikoh2013a}.
However, understanding the large spin lifetimes observed in p-type
Si at room temperature remains an open question \cite{Shikoh2013}.

\begin{figure*}[t]
\begin{centering}
\includegraphics{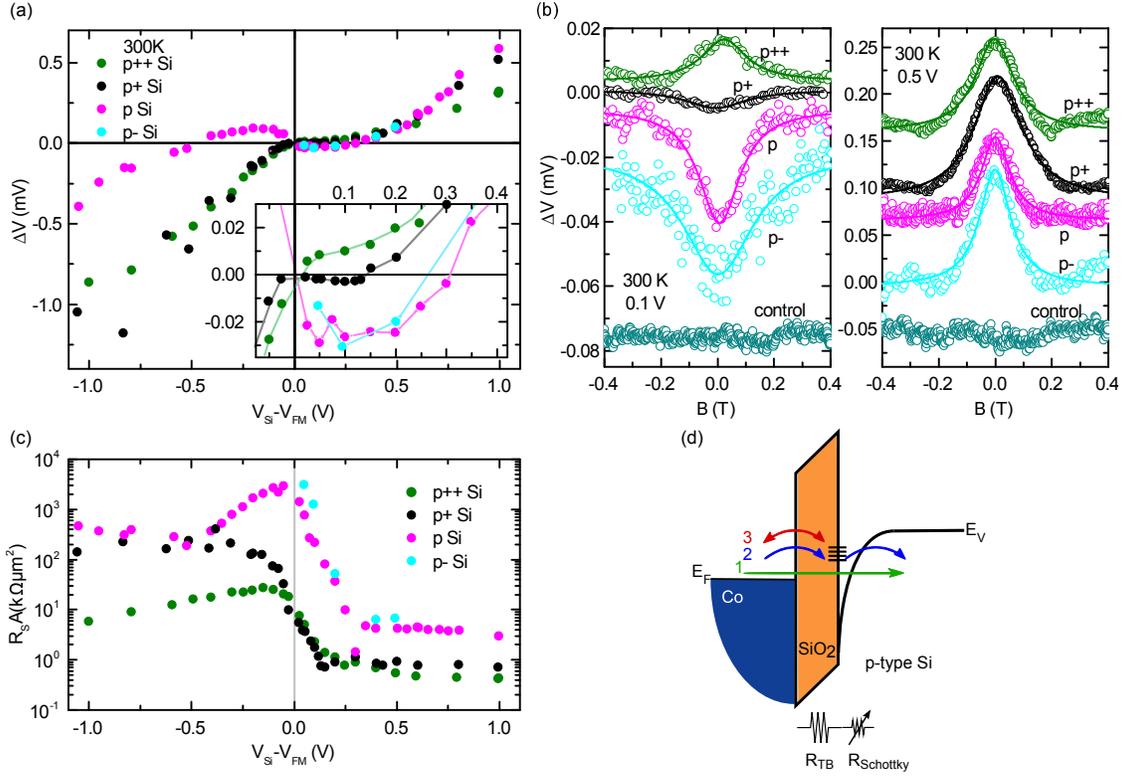} 
\end{centering}

\caption{\textbf{Bias dependence of spin signal with tailored Schottky barrier
width at room temperature:} (a) Bias dependence of Hanle spin signals
for four different boron doping concentrations in p-type Si. The degenerate
p++ Si device shows normal Hanle signal behavior, whereas the nondegenerate
devices (p+, p and p- Si) show anomalous sign reversal. The inset
shows the low bias regime, in order to emphasize the sign reversal
of the spin signals. (b) Hanle curves at $\unit[100]{mV}$ (left panel)
and $\unit[500]{mV}$ (right panel) demonstrating the sign reversal
exists only in the low bias regime. A control sample with nonmagnetic
layer between FM and tunnel barrier demonstrates clearly the origin
of the spin signal in the Si band. (c) $R_{S}A$ with bias voltage
for four devices with different boron doping concentrations in Si.
(d) Energy-band diagram showing different possible spin-transport
mechanisms across the tunnel junction, depending on the resistance
of the tunnel and Schottky barrier. For a fixed SiO$_{2}$ tunnel-barrier
resistance, a low Schottky barrier resistance leads primarily to direct
tunneling between the ferromagnet and the Si (1). With increasing
Schottky barrier resistance, the two-step tunneling into Si via localized
states (2), and the tunneling between the ferromagnet and the localized
states at the interface (3), become more dominant.}

\label{pTypeSpin} 
\end{figure*}

\noindent \textbf{Spin injection into nondegenerate Si and anomalous
spin signals.} In the previous sections, we have presented results
on spin injection into degenerate Si with very narrow Schottky barriers,
where the transport process is dominated by direct tunneling. However,
in nondegenerate Si, the presence of a wider Schottky barrier alters
the transport process. In order to determine the effects of the Schottky
barrier profile on the spin injection and detection process, we perform
experiments on Si with different doping densities, i.e. with different
Schottky barrier widths $W$ and hence resistances (see Fig. \ref{pTypeIntro}a).
We use p-type Si with four different boron doping concentrations (see
Table \ref{SiParameters} for the parameters of the Si samples). The
preparation conditions of the ozone-oxidized SiO$_{2}$ tunnel barrier
and ferromagnetic contacts are kept identical for all these devices.
Figure \ref{pTypeIntro}b shows the significant decrease in current
density or increase in resistance in the reverse bias regime upon
lowering the boron doping density, showing the transition of the contacts
from tunneling to diodic behavior. Such systematic variation of the
Schottky barrier resistance is quite useful in verifying the proposed
spin-transport models.

Hanle spin signals with a Lorentzian line shape have been successfully
measured for all four p-type Si samples at room temperature (Fig.
\ref{pTypeIntro}c). The magnitude of the spin signal is found to
increase with decreasing doping density or with increasing resistance
of the Schottky barrier. Such scaling of the spin signal has been
recently demonstrated with tunnel barrier resistance\cite{Spiesser2012}.
We do not observe a change in the effective spin life time with decreasing
doping concentration in silicon, which is in agreement with the lower
limit for the derived spin lifetime $\tau_{eff}$ \cite{Dash2011a}.

Studying the bias dependence of the Hanle spin signals is an excellent
way to investigate the role of the Schottky barrier, as the applied
bias voltage defines the energy profile at which both spin injection
and detection take place. Figure \ref{pTypeSpin}a shows this bias
dependence of the Hanle signals on the different boron doped Si devices.
With decreasing doping concentration, the width of the Schottky barrier
increased, leading to increased resistance, which in turn yielded
an unusual sign change in the Hanle signal at low bias voltages. 
\begin{table}[h]
\begin{centering}
\begin{tabular}{cccccc}
\hline 
\multicolumn{1}{c}{Si type} & Dopant  & Doping  & Resistivity  & Mobility  & Schottky\tabularnewline
 &  & density  &  &  & barrier width\tabularnewline
 &  & (cm$^{-3}$)  & ($\Omega\,\text{cm}$)  & ($\frac{\text{cm}^{2}}{\text{V\,\ s}}$)  & (nm)\tabularnewline
\hline 
n++  & As  & 3$\cdot10^{19}$  & 0.003  & 118  & 3\tabularnewline
p++  & B  & 1.8$\cdot10^{19}$  & 0.005  & 51  & 7\tabularnewline
p+  & B  & 1.5$\cdot10^{19}$  & 0.008  & 57  & 8.3\tabularnewline
p  & B  & 5.4$\cdot10^{18}$  & 0.011  & 109  & 13.5\tabularnewline
p-  & B  & 1.3$\cdot10^{15}$  & 10  & 466  & 736\tabularnewline
\hline 
\end{tabular}
\end{centering}

\caption{Parameters for degenerate n-type Si and the four p-type Si samples
studied.}

\label{SiParameters} 
\end{table}

For nondegenerate p+ Si, an unusual sign reversal in the spin signal
is detected in the spin extraction regime ($0<V_{Si}-V_{FM}<\unit[140]{mV}$).
By further decreasing the doping concentration (p Si), the sign reversal
of the Hanle signal becomes more prominent, that is, it increases
in magnitude and extends to the higher bias regime of $\unit[400]{mV}$.
Interestingly, a sign change is also observed in the spin injection
regime up to bias voltages of $\unit[-600]{mV}$. Finally, for the
lowest doped p- Si, the sign-reversal phenomenon is also observed
in the spin extraction regime. In this case, the spin injection signal
in the reverse bias condition could not be measured, as the higher
resistance of the Schottky barrier decreased the signal-to-noise ratio.
It should be noted that the sign reversal behavior of the Hanle signal
is observed only at lower bias voltages; at higher bias voltages,
the expected sign of the spin signal is restored. This is demonstrated
exemplary in Fig. \ref{pTypeSpin}b for all four different p-type
Si devices at two different bias voltages: At a low bias voltage of
$\unit[0.1]{V}$ a clear spin signal sign change was observed, whereas
at a higher bias voltage of $\unit[0.5]{V}$ regular Hanle signals
without any sign inversion could be obtained. The detailed measurement
of Hanle signals at different bias voltages, showing sign change for
different p-type Si devices, are presented in supplementary Fig. S3
to Fig. S6. The sign reversal of the Hanle spin signal is also reproduced
for different SiO$_{2}$ tunnel barrier thicknesses and with an Al$_{2}$O$_{3}$
tunnel barrier (supplementary Fig. S7 and Fig. S8). To confirm the
spin-polarized tunneling as origin of the Hanle signals, a control
sample, prepared with a nonmagnetic interlayer ($\unit[10]{nm}$ Ti)
between the Co and the SiO$_{2}$ tunnel barrier, is measured resulting
in no spin signal (Fig. \ref{pTypeSpin}b and supplementary Fig. S10)\cite{Patel2009}.
The $R_{S}A$ values measured on the four doping concentrations are
shown in Fig. \ref{pTypeSpin}c. In addition to the asymmetric behavior
of the signal, a peak also appears at low bias voltages for p Si and
p- Si devices. This enhancement of the spin signal specifically occurs
in the bias voltage range where a sign inversion in the Hanle signal
has been observed.

\part*{Discussion}

In this report we have presented results of large spin accumulations
in degenerate Si by using an ozone oxidized SiO$_{2}$ tunnel barrier.
Narrow Schottky barriers in case of highly doped Si and an improved
interface quality of the SiO$_{2}$ tunnel barrier allows dominant
direct tunneling of spin polarized electrons. The observed large magnitude
of the spin accumulation can not be explained by the standard spin
diffusion model \cite{Fert2001}. Such large spin $R_{S}A$ has also
been observed in literature and reviewed recently in detail \cite{Jansen2012a,Jansen2012c}.
Previously, various control experiments could rule out any enhancement
of the spin signal at room temperature \cite{Dash2009,Jansen2007},
whereas a strong enhancement at low temperature was attributed to
spin accumulations in localized states \cite{Jain2012b,Tran2009}.
Here, we observe a very low temperature dependence with SiO$_{2}$
tunnel barrier which rules out any signal enhancement over the range
from $\unit[5-300]{K}$. This is further supported by the absence
of any variation of spin lifetime with temperature and bias voltage
\cite{Jain2012b,Tran2009}. This demonstrates that high quality interfaces
made of ozone oxidized SiO$_{2}$ on degenerate Si, for a narrow Schottky
barrier, allows efficient spin injection by direct tunneling mechanism.
Nevertheless, a unified theory to explain large spin accumulation
observed in semiconductors is still missing.

Furthermore, our experiments systematically show that larger Schottky
barrier resistances induce and enhance the sign inversion of the spin
signal. Such wider Schottky barrier decouple the localized states
at the interfaces from the silicon bulk bands \cite{Jansen2012b}.
The anomalous spin-signal signs observed in different bias regimes
can be due to competing transport processes across the tunnel and
Schottky barrier \cite{Jansen2012b}. There are three main transport
processes, which can contributing in the different regimes (Fig. \ref{pTypeSpin}d):
(1) At low Schottky barrier resistance (degenerate Si), or at high
bias voltages, direct tunneling dominates, yielding a normal sign
for the Hanle signal. (2) Resonant tunneling via localized states
at the interface can occur for higher Schottky barrier resistances
and competes with direct tunneling. At lower bias voltages, different
escape times for up- and down-spins can give rise to an opposite spin
accumulation resulting in an inversion of the Hanle signal. (3) The
tunneling between the ferromagnet and the localized states can be
dominant when the Schottky barrier resistance is very high, or when
the applied bias voltage is low. 

Previously, several experimental observation of bias dependent sign
inversions of the spin signal have been made in both nonlocal and
three-terminal measurement geometries \cite{Lou2007,Salis2011,Jeon2012b}.
Such sign inversion can be due to: (a) The energy at which the injection
and detection takes place can be different giving rise to different
and even negative spin polarization values for injection and detection
\cite{Sharma1999}. (b) Different escape times of the spin carriers
from localized states can give rise to spin accumulation with opposite
spin orientation \cite{Dery2007a}. (c) The presence of acceptor and
donor states of paramagnetic centers at the interfaces. Those mechanisms
combined with complicated transport processes across the tunnel and
Schottky barrier could be responsible for the sign reversal of the
spin signal. However, this behavior is not clearly understood for
a three-terminal measurement, since the Hanle signal arises from the
spin potential difference created by a spin polarized current through
a single tunnel contact. Although, the exact origin of the sign inversion
of the spin signal is not clear yet, our experiments demonstrate that
an increased Schottky barrier resistance can cause these anomalous
behavior at low bias voltages. 

In conclusion, we have demonstrated a giant spin accumulation in highly
doped Si at room temperature using ozone-oxidized SiO$_{2}$/Co tunnel
junctions. We achieved a spin polarization of 34\% in n-type Si, and
10\% in p-type Si. Temperature and bias dependence measurements indicate
that the spin polarization created in highly doped Si is dominated
by a direct tunneling mechanism. The obtained spin polarization in
n-type Si is almost one order of magnitude larger than previously
reported, and is close to the tunnel spin polarization of the ferromagnetic
tunnel contacts. The low-resistivity and defect density of the ozone-oxidized
SiO$_{2}$ tunnel barrier offers a great advantage over other tunnel
barriers, suggesting a new route to producing large spin polarization
in Si at room temperature. Additionally, spintronic devices based
on SiO$_{2}$ tunnel barriers are compatible with and can easily be
integrated into the present Si technology. Opportunities are growing
with the availability of new Heusler alloy materials that possess
larger spin polarizations, and which can also be integrated with Si
\cite{Kimura2012}. Furthermore, we address here the role of the Schottky
barrier in spin injection and detection processes. Schottky barrier
resistances above a critical limit result in a decoupling of spins
in the localized states from the Si bulk bands resulting in an anomalous
sign reversal in the spin signal in the lower bias voltage regime.
This may be due to the domination of two-step tunneling and thermionic
spin transport across the interface. These findings encourages further
investigation to understand the effect of localized states and paramagnetic
centers on the spin signal. This also requires generating an improved
theoretical description of the spin injection and detection process.
These results will enable utilization of the spin degree of freedom
in complementary Si devices and its further development.

\part*{Methods}

\textbf{Device fabrication:} SiO$_{2}$ tunnel junction preparation:
Spin-transport devices with ferromagnetic tunnel contacts of Co/SiO$_{2}$
were fabricated on a number of Si substrates. The substrates were
cleaned, and the native oxide was etched with diluted hydrofluoric
acid. SiO$_{2}$ tunnel oxide was formed by ozone oxidation at room
temperature, with an O$_{2}$ flow of ($\unit[1]{\frac{l}{min}}$)
for $\unit[30]{min.}$ in the presence of a UV radiation source positioned
4 mm above the chips. The O$_{2}$ molecules are broken down to atomic
oxygen by the UV radiation and combine with O$_{2}$ molecules to
form ozone. Ozone oxidation is known to provide ultra-thin, uniform
tunnel barriers, a fact also confirmed by the temperature and thickness
dependence of our junction resistance measurements. The samples were
then transferred to an electron beam deposition system, where $\unit[15]{nm}$
of ferromagnetic Co layer and $\unit[10]{nm}$ of Au capping layer
were deposited. The ferromagnetic contacts were patterned by photolithography
and Ar\textendash{}ion beam etching. Tunnel contact areas of $\unit[200\times100]{\mu m^{2}}$
were used for measurements, allowing a sufficiently good signal-to-noise
ratio. The Cr ($\unit[10]{nm}$)/Au ($\unit[100]{nm}$) reference
contacts on the Si, and contact pads on the ferromagnetic tunnel contacts
were prepared with the use of photolithography and the lift-off method.
The reference Cr/Au contacts were used to source a DC current and
to detect a voltage signal with respect to the ferromagnetic tunnel
contact.

\textbf{Measurements:} The measurements were performed in three-terminal
geometry, with the spin polarization injected and detected using the
same Si/SiO$_{2}$/Co interface (Fig. \ref{nppSi}a). A constant DC
current was sent through this ferromagnetic tunnel interface into
the Si, and the resulting voltage detected using a nanovoltmeter.
The SiO$_{2}$ tunnel resistance is much higher than that of the Co/Au
metal electrodes and the degenerate Si; for nondegenerate Si, however,
the Schottky barrier resistance dominates and determines the voltage
drop. Measurements were performed in a variable temperature cryostat
($\unit[5]{K}$\textendash{}$\unit[300]{K}$) with a superconducting
magnet. Initially, a sufficiently large in-plane magnetic field (parallel
to the tunnel interface) was applied to create a homogeneous in-plane
magnetization of the ferromagnetic electrode. The spins injected into
the Si by the source current were thus also oriented along the same
axis. In order to measure the resulting spin accumulation in Si, we
employed the Hanle effect, in which a small out-of-plane magnetic
field is applied perpendicular to the spin direction. This has no
significant effect on the orientation of the magnetization of the
ferromagnetic electrode, but it does lead to precession of the spins
in the Si, and thereby a controlled suppression of the spin accumulation.
This changes the tunnel resistance of the active contact, which in
turn produces a change in the detected voltage proportional to the
spin accumulation.

$\vphantom{1cm}$

\part*{Acknowledgment}

The authors acknowledge the support of colleagues at the Quantum Device
Physics Laboratory and Nanofabrication Laboratory at Chalmers University
of Technology. The authors would also like to acknowledge the financial
supported from the Nano Area of the Advance program at Chalmers University
of Technology, EU FP7 Marie Curie Career Integration grant and the
Swedish Research Council (VR) Young Researchers Grant.

\part*{Author contribution}

A.D. fabricated and measured most of the devices. S.P.D and R.S.D
contributed in some fabrication and measurements. S.P.D. and A.D.
conceived the ideas and designed the experiments. S.P.D. planned and
supervised the research. All authors contributed in the discussion
and analysis of the measurements. S.P.D and A.D wrote the manuscript
with input from R.S.D.

\part*{Additional information}

\textbf{Competing financial interests: }The authors declare no competing
financial interests.


\end{document}